\documentclass[prb,aps,twocolumn]{revtex4-2}
\usepackage{natbib}
\usepackage[english]{babel}
\usepackage[T1]{fontenc}
\usepackage{graphicx}
\usepackage{physics}
\usepackage{amssymb}
\usepackage{amsthm}
\usepackage[version=4]{mhchem}
\usepackage[table]{xcolor}
\usepackage{soul}
\usepackage{siunitx}
\usepackage{lipsum}  
\usepackage[backref=true,colorlinks=true,citecolor=blue,urlcolor=blue,linkcolor=blue]{hyperref}

\DeclareSIUnit\angstrom{\text{\AA}}

\newcommand*{\rhoyx}{\rho_{yx}}
\newcommand*{\rhoxx}{\rho_{xx}}

\begin{document}
\sloppy

\title{Doping dependence of the nonlinear Hall resistivity in electron-doped Pr$_{2-x}$Ce$_{x}$CuO$_{4 \pm \delta}$}

\author{M. Dion}\thanks{These authors contributed equally to this work.}
\author{S. Ghotb}\thanks{These authors contributed equally to this work.}
\author{G. Hardy}
\author{P. Fournier}\email[Corresponding author: ]{patrick.fournier@usherbrooke.ca}
\affiliation{Institut quantique, Regroupement québécois sur les matériaux de pointe et Département de physique, Université de Sherbrooke, Sherbrooke, J1K 2R1, Québec, Canada}

\begin{abstract}
    We report on a systematic study of the field dependence of the Hall resistivity $\rhoyx (B)$ as a function of doping in thin films of electron-doped superconducting cuprate $\ce{Pr_{2-x}Ce_{x}CuO_{4 \pm \delta}}$ (\ce{PCCO}). Across the studied doping range from $x = 0.125$ to $x = 0.20$, we observe a nonlinear dependence of $\rho_{yx}$ with $B$. The leading $B^{3}$ nonlinear term is negative, increases with decreasing temperature and peaks  around optimal doping ($x = 0.15$). The observed nonlinear contribution is consistent with the presence of two different types of free carriers (electron-like and hole-like) even for doping with only an apparent hole Fermi surface as observed by angle-resolved photoemission spectroscopy. Based on an analysis using the two-carrier model, this negative nonlinear contribution to $\rhoyx (B)$ implies that the density of the charge carriers behaving like electrons is larger than the density of those behaving like holes for all the doping values explored. Combined with the Hall coefficient reaching $R_{\mathrm{H}}=0$ at specific temperatures for selected doping levels, we also conclude that the mobility of hole-like carriers is larger than that of electron-like carriers for doping around $x^{*}  \approx 0.165$.
\end{abstract}

\maketitle

\section{Introduction}

Materials with strong electronic correlations are routinely studied using the Hall effect as a sensitive probe of the transformations of their electronic states in proximity to the Fermi level. For instance, it is particularly sensitive to the presence of the pseudogap in superconducting cuprates, a partial depletion of the electronic density of states at the Fermi level responsible for a Fermi surface (FS) reconstruction~\cite{hussey2018tale,matsui2007evolution}. For hole-doped cuprates, the zero-temperature Hall effect~\cite{balakirev2003signature} can be correlated to the Fermi surface observed using angle-resolved photoemission spectroscopy (ARPES)~\cite{damascelli2003angle,doiron2017pseudogap}. Most notably, a sharp change in the doping dependence of the positive zero-temperature Hall coefficient, $R_\mathrm{H}(T=0)$, at $x_{\mathrm{FSR}} \sim 0.18 - 0.23$ in hole-doped cuprates signals the closing of the pseudogap as the Fermi surface transitions from nodal arcs at low doping to a large hole-like cylinder as observed by ARPES~\cite{proust2019remarkable,doiron2017pseudogap}. $R_{\mathrm{H}}$ changes even sign at extremely large doping in \ce{La_{2-x}Sr_xCuO_{4}}~\cite{tsukada2006negative} as one would expect from a band filling perspective when the hole doping level is such that the FS becomes electron-like. Despite the closing of the pseudogap at $x_{\mathrm{FSR}}$, the Hall coefficient of cuprates remains strongly temperature dependent for all doping~\cite{ando2004evolution,smith1910variation,ong1990hall}, except perhaps for the largest doping reported in Ref. \onlinecite{tsukada2006negative}. This strong temperature dependence of $R_{\mathrm{H}}$ remains one of the biggest theoretical challenges in the understanding of the electronic properties of cuprates. One attempt to explain such temperature dependence was introduced for example by Kontani \textit{et al.} taking into account the strong effect of temperature-dependent antiferromagnetic fluctuations (AF) leading to current vertex corrections (CVC) for a nearly antiferromagnetic Fermi liquid (NAFL)~\cite{kontani1999hall,jenkins2010origin}. It describes how the dynamics of electrons in the presence of crossed electric and magnetic fields
is affected beyond the impacts of the changes imposed to the band structure by strong interactions as probed for instance by ARPES.

The evolution with doping of the Fermi surface of the electron-doped cuprates \ce{R_{2-x}Ce_{x}CuO_{4}} (R = La, Pr, Nd, Sm) as observed by ARPES~\cite{matsui2007evolution,dagan2004evidence,armitage2002doping} is somewhat different from hole-doped cuprates but still correlates with the doping dependence of the zero-temperature $R_{\mathrm{H}}$~\cite{dagan2004evidence,charpentier2010antiferromagnetic,armitage2010progress}. As a distinctive feature, only anti-nodal arcs are first observed at low doping while the Hall coefficient is fully negative~\cite{charpentier2010antiferromagnetic,dagan2004evidence,matsui2007evolution}. At intermediate doping, the emergence of nodal arcs leads to hole-like carriers and a change of sign of $R_{\mathrm{H}}(T=0)$ close to optimal doping as the FS completes a reconstruction at $x^{*} \sim 0.165$ for \ce{Pr_{2-x}Ce_{x}CuO_{4}} from arcs to a large hole-like cylinder similar to the hole-doped one~\cite{dagan2004evidence,charpentier2010antiferromagnetic}. The doping dependence of $R_{\mathrm{H}}(T=0)$ and ARPES is consistent with reports that transport properties can be qualitatively explained using a two-carrier model~\cite{matsui2007evolution,armitage2002doping,li2007normal,dagan2004evidence,fournier1997thermomagnetic}.

It remains however a challenge to fully explain the temperature dependence of the Hall coefficient of electron-doped cuprates~\cite{armitage2010progress}. For instance, even though the FS has fully reconstructed into a large hole-like cylinder for doping just above $x^* \sim 0.165$, this Hall coefficient for $x$ ranging from $0.17$ to $\sim 0.19$ changes sign as temperature is varied~\cite{armitage2010progress,charpentier2010antiferromagnetic,dagan2004evidence}. From the low temperature positive value, it becomes negative at intermediate temperature, clearly demonstrating that electronic states on portions of this large hole-like FS are still behaving like electrons, thus contradicting the ARPES results that indicate no change in FS shape. This characteristic sign change is however compatible with the theory mentioned above proposed by Kontani \textit{et al.}~\cite{kontani1999hall} to explain the temperature dependence of $R_{\mathrm{H}}$ for the hole-doped cuprates. As transport properties remains qualitatively consistent with the expectations of a two-carrier model even for doping beyond $x^{*}$, it becomes interesting to check if the analysis of the temperature dependence of the resistivity and the Hall effect can be pushed further with this simple model, with parameters that could eventually be linked to the parameters of a theory taking into account the impact of strong electron interactions, for example as was introduced in Ref. \cite{kontani1999hall}.

In a recent paper, we show that the Hall resistivity $\rhoyx $ of overdoped electron-doped \ce{Pr_{1.82}Ce_{0.18}CuO_{4}} (\ce{PCCO}) presents a nonlinear field dependence at moderate magnetic fields $B$~\cite{ghotb2024}.
In the limit of small applied magnetic field, the field dependence of the Hall resistivity in the two-carrier model can be expressed as an expansion over odd powers of $B$,
\begin{align}
    \rhoyx  (B) = R_{\mathrm{H}} B+C B^{3}+D B^{5} + ...
    \label{eq:res_hall_expansion}
    .
\end{align}
In particular, in the vicinity of the temperature, $T_{\mathrm{cr}}$, where $R_{\mathrm{H}}(T)$ is zero, the leading nonlinear $B^{3}$ contribution can be easily observed when measuring $\rhoyx (B)$. The observed negative value of $C$ is used to demonstrate that the density of the carriers behaving like electrons in PCCO with $x = 0.18$ is larger than the density of the hole-like carriers using the two-carrier model. Combining that with $R_{\mathrm{H}}=0$ at $T_{cr}$ allows us also to conclude that the hole mobility is larger than the electrons' one~\cite{ghotb2024}. In fact,  nonlinear contribution can be tracked for temperatures away from $T_{\mathrm{cr}}$ even if the linear term dominates $\rhoyx  (B)$. The nonlinear contributions grow with temperatures. Obviously, it becomes interesting to track the doping dependence of these nonlinear contributions ($C$, $D$, ...) to the Hall resistivity and how they correlate with the evolution of the Fermi surface with doping.

In this paper, we present a systematic study of the Hall resistivity of \ce{Pr_{2-x}Ce_{x}CuO_{4}} thin films focusing on the extraction of the leading nonlinear $B^{3}$ term of its field dependence and tracking its doping dependence. We show that this $B^{3}$ term is non-zero and remains negative for the whole doping range explored from $x = 0.125$ to $0.20$ demonstrating that the density of electron-like carriers is always larger than the hole-like carriers. Moreover, the first nonlinear term ($C$) reaches a maximum absolute value around optimal doping $x \approx 0.15$. This demonstrates the applicability of the two-carrier model at all doping even beyond $x^{*} = 0.165$ corresponding to the critical doping for a Fermi surface reconstruction. We argue that the presence of electron-like carriers for doping values beyond $x^{*}=0.165$ can be explained by a theoretical proposal that electron-like behaviour of carriers on a hole-like Fermi surface can arise from antiferromagnetic fluctuations.

\section{Experiments and methods}

Thin films of $c$-axis oriented $\ce{Pr_{2-x}Ce_{x}CuO_{4 \pm \delta}}$ with various \ce{Ce} concentrations ($0.05 \leq x \leq 0.20$) are deposited by pulsed laser deposition (PLD) from an off-stoichiometric target with excess \ce{CuO}~\cite{roberge2009improving} on (100)-oriented \ce{La_{0.18}Sr_{0.82}Al_{0.59}Ta_{0.41}O_{3}}  (\ce{LSAT}) substrates using a \SI{248}{\nm} KrF excimer laser. All samples of typical thickness of \SIrange[range-units=single, range-phrase = --]{100}{120}{\nm} are grown from \SIrange[range-units=single]{820}{840}{\degreeCelsius} in a \ce{N_{2}O} atmosphere at a pressure of \SI{200}{mTorr}. To obtain the maximum transition temperature and high crystalline quality, the growth temperature for each Ce concentration is optimized as well as the in-situ post-annealing required to achieve superconductivity. The crystal structure of the layers is characterized using a Bruker AXS D8 Discover X-ray diffractometer (XRD) in the $2 \theta - \omega$ configuration, which also allows to estimate the layers' thickness using a Fourier transform approach~\cite{dion_interface_2017}.

To achieve precise electrical transport measurements and to minimize the impacts of post-growth sample processing, we employ three distinct fabrication methods for Hall bars, which are explained in detail in \cite{ghotb2024}. In brief, we compared Hall bars defined by photolithography in two different fabrication processes: one involving etching of the films using Ar ion milling on a liquid nitorgen cooled stage, and another based on a substrate selective epitaxy technique similar to that explored by Morales \textit{et al} \cite{morales2005selective,ma1989novel}. Finally, we compared these lithography samples to Hall bars defined by hand using a fine diamond scriber tip. Each of these methods yielded consistent resistivity and Hall effect results.
The longitudinal and Hall resistivity measurements, $\rhoxx$ and $\rhoyx$ respectively, are carried out using a standard four-probe method in a Physical Property Measurement System (PPMS) from Quantum Design with a magnetic field applied perpendicular to the sample surface up to \SI{9}{\tesla} in the temperature range from \SIrange[range-units=single]{2}{300}{\kelvin}. For the sensitive Hall voltage measurements required to observe the nonlinear $\rhoyx (B)$, an external current source and nanovoltmeters are used to improve the signal to noise ratio. In this specific case, the Hall (transverse) voltage is measured at a fixed temperature as a function of positive and negative applied magnetic field. Special care is taken to stabilize the field prior to measurements and the Hall resistivity is computed to eliminate any remaining offset due to the misalignment of the Hall contacts. For the proper extraction of $\rhoyx(B)$, the transverse voltage ($V_y$) is anti-symmetrized using positive and negative fields
\begin{align*}
    \rhoyx (B)=\frac{t (V_y(+B)-V_y(-B))}{2I_{x}}
\end{align*}
where $I_x$ is the injected current and $t$ is the sample thickness (along the magnetic field). From Eq.~\ref{eq:res_hall_expansion}, the Hall coefficient $R_{\mathrm{H}}$ is defined as the slope ($\dd{\rho_{yx}}/\dd{B}$) at $B \rightarrow 0$.
Separate measurement runs at fixed magnetic field of $\SI{\pm 9}{\tesla}$ while varying the temperature are used to estimate the temperature dependence of the Hall coefficient. We label the Hall coefficient extracted from these measurement $\tilde{R}_{\mathrm{H}}$ since it is computed as the ratio between the Hall resistivity and the magnetic field $\rhoyx(B)/B$ at $B=\SI{\pm 9}{\tesla}$ and is different from $R_{\mathrm{H}}$ (in Eq.~\ref{eq:res_hall_expansion}) when $\rhoyx(B)$ is nonlinear.


\section{Results and discussion}

\begin{figure}
    \center
    \includegraphics[scale=1]{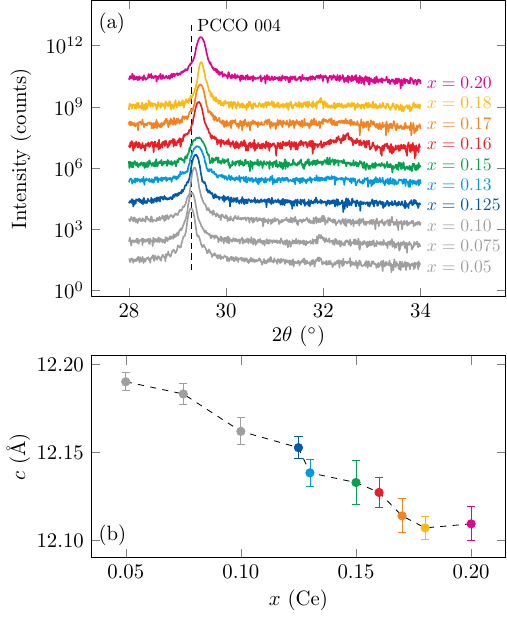}\\
    \caption{(a) X-ray diffraction patterns of the \ce{PCCO} thin films with different \ce{Ce} concentrations in the range of $2\theta =~$\SIrange[range-units=single, range-phrase=--]{28}{34}{\degree}. From bottom to top, the \ce{Ce} concentration increases from \numrange{0.05}{0.20}.  The data has been shifted vertically for clarity. The underdoped region, where the transport properties are not studied, is shown in gray.  (b) The value of the $c$-axis lattice parameter of \ce{PCCO} thin films as a function of \ce{Ce} concentration.}

    \label{fig:xrd_spectrums}
\end{figure}

Figure~\ref{fig:xrd_spectrums}(a) presents the $2 \theta-\omega$ XRD spectra from $2\theta =~$\SIrange[range-units=single]{28}{34}{\degree} of the films in the whole doping range. These spectra focus on the angular range that allows one to detect the (004) diffraction peak of \ce{PCCO} observed just above $2\theta =~$\SI{29}{\degree} and peaks at $\sim~$\SIrange[range-units=single, range-phrase=--]{32}{33}{\degree}, revealing the eventual presence of unwanted parasitic intercalated phases that are known to affect the accurate determination of the absolute value of $\rhoxx$ and $\rhoyx $~\cite{roberge2009improving}. As shown in Fig.~\ref{fig:xrd_spectrums}(a), using off-stoichiometric targets with \SI{5}{\%} excess \ce{CuO} and the optimal growth conditions prevent the formation of these parasitic phases~\cite{roberge2009improving}. The full XRD patterns of the samples indicate that all the PCCO thin films are growing coherently along the $c$-axis with no sign of secondary phases. In Fig.~\ref{fig:xrd_spectrums}(a), the (004) peak of \ce{PCCO} shifts to higher angles with increasing \ce{Ce} concentration, $x$, indicating a decrease in the $c$-axis lattice parameter. Fig.~\ref{fig:xrd_spectrums}(b) shows the value of the room temperature $c$-axis lattice parameter as a function of \ce{Ce} concentration confirming its decrease with increasing \ce{Ce} content. This behavior is consistent with the trend reported before for \ce{PCCO} thin films over the whole doping range~\cite{maiser1998pulsed,armitage2010progress,fournier2015t} resulting from the substitution of \ce{Pr} with an ionic radius of \SI{1.266}{\angstrom} by \ce{Ce} with an ionic radius of \SI{1.11}{\angstrom}. A similar trend has also been observed with \ce{NCCO}~\cite{takagi1989superconductivity,tarascon1989growth}.


\begin{figure}
    \center
    \includegraphics[scale=1]{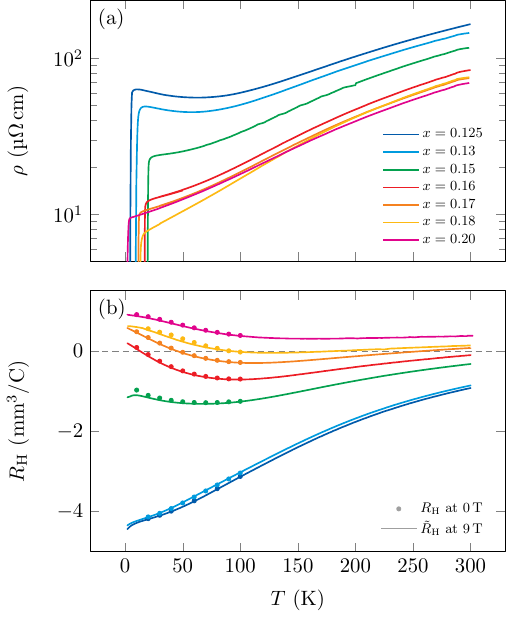}\\
    \caption{(a) Resistivity as a function of temperature of \ce{PCCO} thin films for $0.125\leq x \leq 0.20$; (b) The Hall coefficient as a function of temperature for different \ce{Ce} contents ranging from \numrange{0.125}{0.20}. Full lines are extracted as $\rhoyx(B)/B$ at $B=\SI{\pm 9}{\tesla}$ while dots are the slopes $\dd{\rhoyx}/\dd{B}$ at $B=0$ obtained from the isotherms fits (see Fig.~\ref{fig:rho_yx_fit}).}
    \label{fig:resistivity_and_hall}
\end{figure}

The in-plane resistivity, $\rhoxx$, and the Hall coefficient, $R_{\mathrm{H}}$, as a function of temperature from \SIrange[range-units=single]{2}{300}{\kelvin} for the \ce{PCCO} thin films with $x = 0.125$ to $x = 0.20$ are presented in Figure~\ref{fig:resistivity_and_hall}. The evolution of the transport properties can be divided into three main regions in the phase diagram. In the underdoped region with $x  \lesssim  0.125$ where long-range antiferromagnetic order dominates the phase diagram at low temperature~\cite{armitage2010progress,motoyama2007spin}, the in-plane resistivity displays the usual high-temperature metallic behavior that turns into an insulating-like low-temperature dependence~\cite{fournier1998insulator}. The magnitude of $\rhoxx$ depends strongly on doping changing by at least an order of magnitude. In this same range ($x =~$\numrange{0.05}{0.13}) shown in Ref.~\cite{charpentier2010antiferromagnetic}, the Hall coefficient is large and negative indicating a low density of negative free carriers. $R_{\mathrm{H}}$ at $T = 0$ decreases quickly with $x$ tracking roughly the expected $R_{\mathrm{H}} \sim -1/x$ doping dependence~\cite{charpentier2010antiferromagnetic}. In this doping range, angle-resolved photoemission spectroscopy (ARPES) reveals a single set of arcs close to $(\pm \pi,0)$ and $(0,\pm \pi)$ in the Brillouin zone (often called the anti-nodal positions)\cite{matsui2007evolution,armitage2002doping} and are tied naturally to the negative carriers observed in $R_{\mathrm{H}}$.


\begin{figure*}
    \center
    \includegraphics[scale=1]{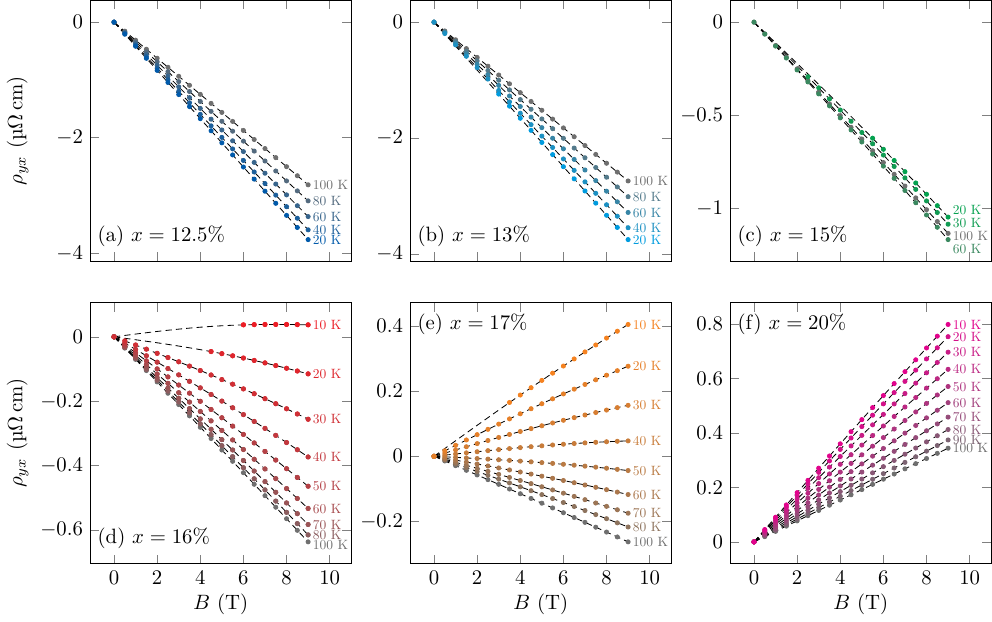}
    \caption{Hall resistivity $\rhoyx $ as a function of magnetic field for $0.125 \leq x \leq 0.20$ at various temperatures above $T_{\mathrm{c}}$. Fits of each isotherm to Eq.~\ref{eq:res_hall_expansion} are shown with dashed lines. Selected isotherms are not shown to avoid overcrowding the figure. }
    \label{fig:rho_yx_fit}
\end{figure*}


Increasing the \ce{Ce} concentration above $x = 0.125$ still leads to a decreasing resistivity, although at a slower rate, and to the emergence of superconductivity. All the films in this doping range have a sharp superconducting transition with a maximum transition temperature of $T_{\mathrm{c}} =~$\SI{21}{\kelvin} at optimal doping ($x = 0.15$). Starting at $x \sim 0.15$, we can observe a low temperature positive upturn to $R_{\mathrm{H}}$ that is building up quickly with increasing doping and leading eventually to a sign change of the zero temperature $R_{\mathrm{H}}$ as previously reported~\cite{dagan2004evidence,charpentier2010antiferromagnetic}. In this doping range, ARPES reveals the anti-nodal arcs previously introduced and a new nodal set with arcs close to $(\pm \pi/2,\pm \pi/2)$ in the Brillouin zone that are associated naturally to the hole-like carriers observed by transport.

In this same doping range, $R_{\mathrm{H}}$ for a specific doping can change sign as a function of temperature. This signature in $R_{\mathrm{H}}(T)$ is central to the present study and can only be explained if both electron-like and hole-like carriers contribute to the current density as was often discussed in previous reports~\cite{armitage2010progress,dagan2005origin,dagan2004evidence,charpentier2010antiferromagnetic,jiang1994anomalous,crusellas1993two,gollnik1998doping}. The crossing temperature, $T_{\mathrm{cr}}$, at which $R_{\mathrm{H}}(T)$ changes sign at low temperature increases with doping. This crossing occurs at \SIlist[list-units=single]{12;44;90}{\kelvin} for $x =~$\numlist{0.16;0.17;0.18}, respectively. A second sign change can be observed at higher temperatures for $x =~$\numlist{0.17;0.18}. Eventually, in the third $x$ range at high enough \ce{Ce} doping (roughly $x \gtrsim 0.19$), the Hall coefficient remains positive over the whole temperature range but preserves a strong temperature dependence. Noticeably, the Hall coefficients extracted at \SI{9}{\tesla} ($\tilde{R}_{\mathrm{H}}$)  and \SI{0}{\tesla} ($R_{\mathrm{H}}$) coincide except at low temperatures and around the optimal doping $x = 0.15$.

Across the entire $x$-$T$ phase diagram, the behavior of the Hall coefficient $R_{\mathrm{H}}$ can be summarized as follows: 1) At low temperatures, the carrier type is consistent with what is observed by ARPES. 2) At high temperatures, the behavior trends towards a hole-like Fermi surface without significant reconstruction occurring at temperatures much higher than those achieved in our experiments. 3) At intermediate temperatures, the system exhibits characteristics of a partitioned Fermi surface, showing contributions from both hole-like and electron-like carriers. Overall, the data presented in Fig.~\ref{fig:resistivity_and_hall} is consistent with previous results on \ce{PCCO}~\cite{dagan2004evidence,charpentier2010antiferromagnetic}.

The persisting electron-like carriers seen in electrical transport for doping above $x^* \sim 0.165$ are in direct contradiction with the single well-defined large hole-like cylindrical Fermi surface observed by ARPES~\cite{armitage2002doping,matsui2007evolution}. In order to generate a negative Hall coefficient, parts of the states on this hole-like Fermi surface must behave like electrons in the presence of crossed applied electric and magnetic fields. As mentioned above, such possibility has been explored for example by Kontani \textit{et al.}~\cite{kontani1999hall} to explain the strong temperature dependence of the Hall effect in cuprates in general, in particular the sign changes observed in $n$-type cuprates as a direct consequence of strong electronic interactions. Applying current vertex corrections (CVC) to a nearly antiferromagnetic Fermi liquid (NAFL), antiferromagnetic fluctuations can lead to contributions to the current density no longer parallel to the velocity vector $\vec{v} (\vec{k})$ (no longer perpendicular to the FS) at specific locations of the FS. In fact, the most important corrections to the current density occur close to the crossing points of the FS with the antiferromagnetic Brillouin zone (AFBZ) limits, i.e. for wavevectors in proximity to the hot spots~\cite{kontani1999hall}. While carriers on the nodal arcs preserve their hole-like character, the anti-nodal ones gain an extra velocity component that transforms them into electron-like excitations at the origin of the negative component in the Hall resistivity.

As a crude model of this resulting partition of the FS into hole-like and electron-like sections, one can use the conventional two-carrier model to capture the overall qualitative behavior of this family. Within this model, the Hall resistivity can be expressed as :
\begin{align}
    \begin{split}
        \rhoyx
         & = \frac{p\mu_p^{2}-n\mu_n^{2}}{e(n\mu_n+p\mu_p)^{2}}B + \frac{np(p-n)\mu_n^{2}\mu_p^{2}(\mu_n+\mu_p)^{2}}{e(n\mu_n+p\mu_p)^{4}}B^{3}
        \\
         & \qquad + \frac{np(n-p)^{3}\mu_n^{4}\mu_p^{4}(\mu_n+\mu_p)^{2}}{e(n\mu_n+p\mu_p)^{6}}  B^{5} +...
    \end{split}
    \label{eq:RH_Eq}
\end{align}
where $n$ and $p$ are the density of electron-like and hole-like carriers, respectively, while $\mu_n$ and $\mu_p$ are their respective mobilities~\cite{ghotb2024}. This equation is obtained as a series expansion assuming that the magnetic field remains small, i.e. for $\mu_n B \ll 1$ and $\mu_p B \ll 1$. In Eq.~\ref{eq:RH_Eq}, we recognize the first term as the usual expression for $R_{\mathrm{H}}$ of the two-carrier model in the low field limit. However, if both $n$ and $p$ are non-zero and of different magnitude, the following terms will be non-zero and result into deviations from linear Hall resistivity. Their sign will reveal also which carrier density is dominating.


Figure~\ref{fig:rho_yx_fit} shows a series of measurements of the Hall resistivity as a function of magnetic field for \ce{PCCO} films for various $x$. The isothermal Hall resistivity is presented for temperatures and magnetic fields outside the superconducting region to avoid the presence of extra field dependence at low field due to superconductivity. The field dependance of $\rhoyx$ is tracked by fitting each isotherm to Eq.~\ref{eq:res_hall_expansion} up to the fifth power of $B$ with very good agreement. We will later make use of the values of $R_\text{H}$ and $C$ extracted from these fits.

In accordance with the $R_{\mathrm{H}}(T)$ data for $x < 0.16$ (Fig.~\ref{fig:resistivity_and_hall}(b)) where the Hall resistivity is negative in the entire temperature range, the isothermal Hall resistivity of this doping range remains negative up to a magnetic field of \SI{9}{\tesla} and its value decreases with temperature. Similarly, a positive isothermal Hall resistivity is observed in the heavily overdoped side ($x=0.20$) for all temperatures. At intermediate doping, the low temperature sign change can also be identified easily in the isothermal Hall resistivity for $x =~$ \numrange{0.16}{0.18}. For instance, the film with $x = 0.17$ in Fig.~\ref{fig:rho_yx_fit}(e) shows a positive $\rhoyx (B)$ at low temperature up to \SIrange[range-units=single]{40}{50}{\kelvin}, then becomes negative above \SI{50}{\kelvin}. It remains negative up to \SI{260}{\kelvin} where it switches back to positive $\rhoyx (B)$ (not shown here). More important, a close inspection of the isotherms at \SIrange[range-units=single]{40}{50}{\kelvin} in proximity to its $T_{\mathrm{cr}} \sim~$\SI{46}{\kelvin} for $x = 0.17$ reveals a deviation from linearity.

\begin{figure}
    \center
    \includegraphics[scale=1]{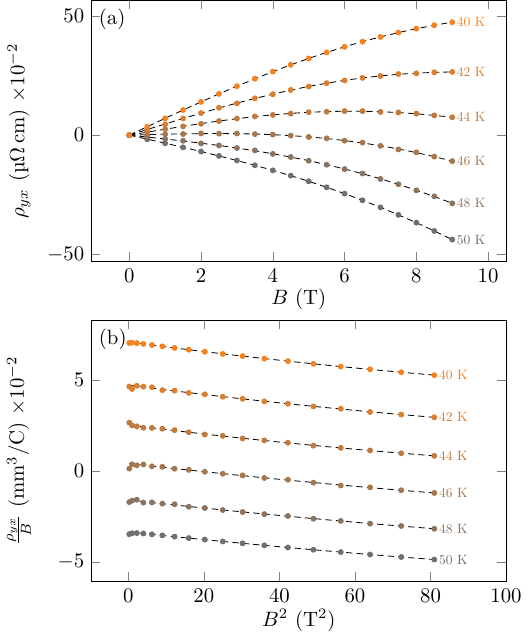}\\
    \caption{(a) Isothermal Hall resistivity as a function of magnetic field at temperature increments of \SI{1}{\kelvin} in the vicinity of the low-T crossing point at $T_{\mathrm{cr1}} \cong~$\SI{44}{\kelvin} for $x = 0.17$. (b) $\rhoyx /B$ as a function of $B^{2}$ for the same dataset.}
    \label{fig:rho_yx_pcco17}
\end{figure}

Figure~\ref{fig:rho_yx_pcco17}(a) shows a series of isotherms measured with a small temperature interval for $x = 0.17$ in proximity of $T_{\mathrm{cr}}$. The downward bending clearly seen here is arising from the higher order terms in the field dependence of the Hall resistivity. In fact, from this plot, one can determine that $R_{\mathrm{H}} = 0$ at $T_{\mathrm{cr}} =~$\SI{46}{\kelvin} (slope is really zero at $B \rightarrow 0$).


In a previous report, we explored this nonlinear term for the specific doping of $x = 0.18$, tracked its temperature dependence and made a connection to the two-carrier model~\cite{ghotb2024}. In order to clearly confirm the presence of these higher order terms with the leading term following $B^{3}$ even for temperatures where it is barely observable as it is overwhelmed by the large linear term, we can plot the Hall resistivity data as $\rhoyx /B$ as a function of $B^{2}$ as shown in Fig.~\ref{fig:rho_yx_pcco17}(b) for the film with $x = 0.17$ at temperatures close $T_{\mathrm{cr}}$. On this plot, the first nonlinear contribution can be observed as the slope close to $B \sim 0$.


\begin{figure*}
    \center
    \includegraphics[scale=1]{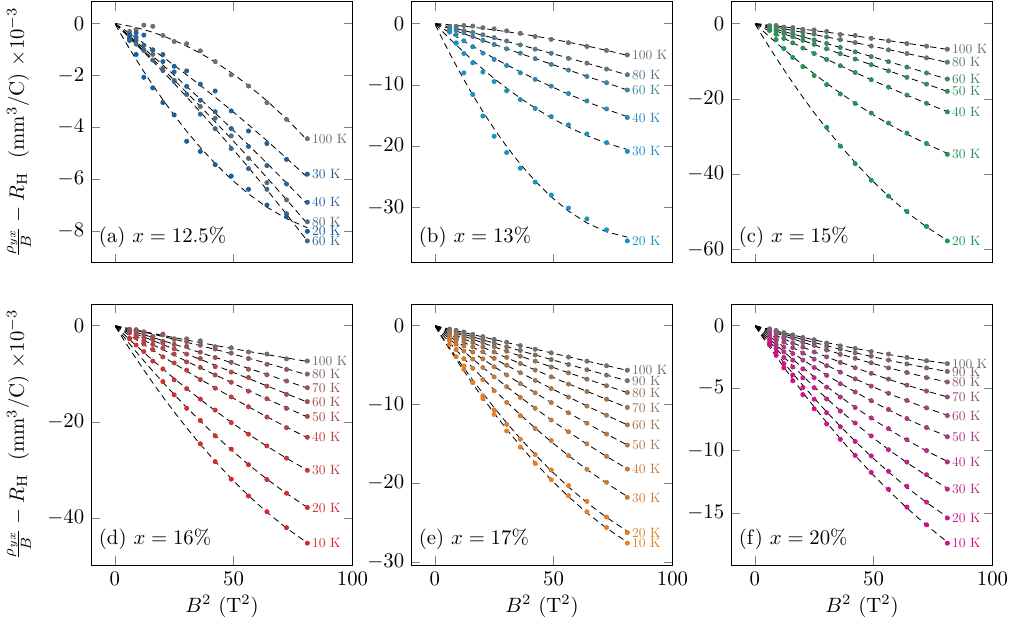}
    \caption{Hall resistivity $\rhoyx /B - R_{\mathrm{H}}$ as a function of $B^{2}$ for $0.125 \leq x \leq 0.20$, extracted from the data in Fig.~\ref{fig:rho_yx_fit}. Data for $B<\SI{2}{\tesla}$ are not shown.}
    \label{fig:rho_yx_over_field_shift_fit}
\end{figure*}


Fig.~\ref{fig:rho_yx_pcco17}(b) reveals the change in sign of $R_{\mathrm{H}}$ (intercepts at $B^{2} \rightarrow 0$) around $T_{\mathrm{cr}}$, but most importantly the presence of a negative $C$ persisting for the temperatures close to $T_{\mathrm{cr}}$ even when $R_{\mathrm{H}} \neq 0$. According to Eq.~\ref{eq:RH_Eq}, a negative $C$ is only possible if $n > p$~\cite{ghotb2024}. Focussing on $x$ values above $x^* = 0.165$, this implies that large portions of the hole-like FS seen by ARPES~\cite{matsui2007evolution} correspond to states where carriers behave like electrons despite the hole-like FS. Using the fact that $R_{\mathrm{H}} = 0$ at $T_{\mathrm{cr}}$ leads to $n \mu_n^{2} = p \mu_p^{2}$ in the leading term $R_{\mathrm{H}}$ of Eq.~\ref{eq:RH_Eq}, and since $n > p$ from $C < 0$, we deduce that $\mu_n < \mu_p$ at this specific temperature for $x = 0.17$. In electron-doped cuprates, although the density of hole-like carriers is smaller than the density of electron-like carriers, their mobility is larger than that of the electrons and they become the main contribution to the low temperature conductivity and lead to positive Hall effect.

As was noticed in Ref.~\cite{ghotb2024}, the higher order terms ($B^{3}$ and beyond) can be observed at all temperatures. Even though it cannot be easily noticed in the data in Fig.~\ref{fig:rho_yx_fit} because the leading term $R_{\mathrm{H}} B$ dominates the signal, using the plots of $\rhoyx /B - R_{\mathrm{H}}$ as a function of $B^{2}$ allows us to isolate the nonlinear contributions to the Hall resistivity as a function of temperature for all $x$ values. Figure~\ref{fig:rho_yx_over_field_shift_fit} shows the same data as Fig.~\ref{fig:rho_yx_fit} presented as $\rhoyx /B - R_{\mathrm{H}}$ as a function of $B^{2}$ for $0.125 \leq x \leq 0.20$. On these plots, the slope around $B\sim 0$ is directly equal to $C$. We notice that it is present and negative for all dopings in the range $0.125 \leq x \leq 0.20$ and that it is increasing with decreasing temperatures. At the lowest temperatures, we can detect additional contributions to the nonlinear Hall resistivity that are associated to the $B^5$ and following terms in Eq.~\ref{eq:RH_Eq}.

\begin{figure}
    \center
    \includegraphics{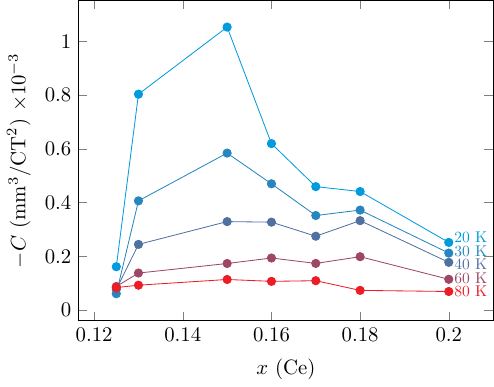}\\
    \caption{The coefficient of the nonlinear $B^{3}$ term of the Hall resistivity as a function of \ce{Ce} concentration for ($0.125 \leq x \leq 0.20$) obtained from the slope and the intercept of the plot in Fig.~\ref{fig:rho_yx_over_field_shift_fit}. }
    \label{fig:rho_yx_c_coefs}
\end{figure}

Figure~\ref{fig:rho_yx_c_coefs} presents the doping dependence of $C$ for $0.125 \leq x \leq 0.20$ in the temperature range of \SIrange[range-units=single]{30}{90}{\kelvin} obtained from the data in Fig.~\ref{fig:rho_yx_over_field_shift_fit}. The first significant observation is that $C$ remains negative over the whole doping range of interest. The sign is not affected by the Fermi surface reconstruction occurring at $x^{*} = 0.165$ where the pseudogap closes at the hot spots and the Fermi surface changes from arcs to a full hole-like cylinder~\cite{armitage2002doping,matsui2007evolution}. This indicates that the carrier signs on portions of the FS that merge at the pseudogap closing line at $T^* (x)$~\cite{zimmers2005infrared} leading to the large hole-like cylinder do not change abruptly with doping around $x^{*}$. We notice also that $C$ decreases rapidly from $x=0.13$ to $x=0.125$. Unfortunately, the evaluation of $C$ for $x < 0.125$ becomes difficult as weak localization effects with large magnetoresistance accompanied by large field dependence of the Hall resistivity~\cite{fournier2000anomalous,armitage2010progress,greene2020strange,rullier2001disorder} blurs the exploration of the nonlinear Hall resistivity. It then becomes difficult to confirm its full disappearance or even its sign change in the same range of doping where the nodal arcs seem to disappear according to ARPES~\cite{armitage2002doping,matsui2007evolution}.

\begin{figure}
    \center
    \includegraphics{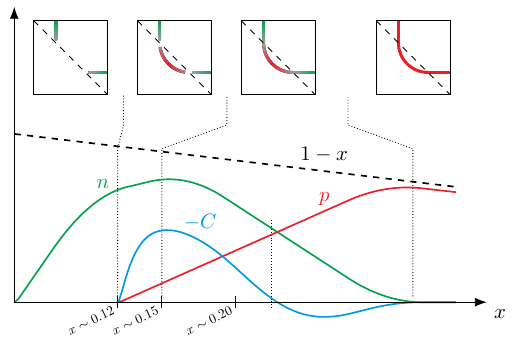}\\
    \caption{A sketch of a possible scenario within the two-carrier model to explain the doping dependence of $C$. It shows the approximate doping dependence of $n$ and $p$, the density of electron-like (solid green line) and hole-like (solid red line) carriers, and a sketch of the expected value of $-C$ the second leading term of the Hall resistivity (solid blue line). Upper panels illustrate the Fermi surface evolution as is observed by ARPES. Carriers on portions in red behave as holes while they behave as electrons in green. Gray portions are meant to show the zones of strongest influence of strong electron correlations as proposed in Ref.~\cite{kontani1999hall}.}
    \label{fig:sketch}
\end{figure}

Nevertheless, Fig.~\ref{fig:rho_yx_c_coefs} clearly shows a change of trend of $C(x)$ around optimal doping and $x^{*}$. Reminding that $C \propto np (p-n)$ from Eq.~\ref{eq:RH_Eq}, it may indicate a change of trend in $n$ and/or $p$ as a function of doping, or more precisely a change of trend in the impact of strong electronic interactions in a more complete theoretical approach. A sketch of a possible scenario for the doping dependence of $C$ is shown in Figure~\ref{fig:sketch}. This scenario is relying on the expectations that strong electron correlations modify mostly the dispersion and the electronic dynamics in the vicinity of the hot spots, the crossing points of this large hole cylinder with the AFBZ boundaries. Coming from the high doping side where a full hole-like FS leads to a positive $R_{\mathrm{H}} \sim 1/p$ where $p \sim 1-x$ with $n \sim 0$ and $C$ small, a gradual increase of the portion of the FS behaving like electrons rather than holes leads to a decreasing $p$ (away from 1-$x$) and an increasing $n$ as $x$ decreases. This transformation of the electronic behavior is compatible with the growing influence of  antiferromagnetic fluctuations as proposed by Kontani \textit{et al.}~\cite{kontani1999hall} even if the FS is a large hole cylinder. As $x$ reaches $\sim x^{*}$, $n$ reaches a maximum. Indeed, as the opening of the pseudogap starts removing states from both sides of the AFBZ boundary, it leads likely to a decrease in $n$ and $p$ with further decreasing doping. Altogether, it results in a maximum in the magnitude of $C(x)$ in proximity of $x^{*}$. Assuming finally a stronger suppression of the nodal arcs than the anti-nodal ones as is observed by ARPES~\cite{armitage2002doping,matsui2007evolution}, the hole carrier density vanishes between $x \sim~$\numlist{0.10;0.13} resulting also in a vanishing of $C$ on the underdoped side.


Nonlinear Hall resistivity has already been reported for cuprates. It was reported by Li \textit{et al.}~\cite{li2007high} in \ce{PCCO} thin films in the very high-field regime compared to ours, as large as \SI{58}{\tesla} using pulsed magnetic fields. Their observations of a large positive curvature in $\rhoyx (B)$, combined to the change from a quadratic to a linear field regime in the magnetoresistance, led the authors to propose a spin-density-wave induced FS reconstruction as the possible origin of these high-field manifestations. We are not debating these results and the related interpretation as we are focussing our attention to a much lower magnetic field range than in Ref.~\cite{li2007high} with our maximum applied field of \SI{9}{\tesla} as we intended to stay in the low field limit. The effect we observe is also much weaker than that observed in Ref.~\cite{li2007high} and may not be related directly (not the same origin). Nonlinear Hall resistivity was also reported in hole-doped cuprates and analyzed using the two-carrier model~\cite{rourke2010fermi}. In this particular case, the two types of carriers in this underdoped material are attributed to a FS reconstruction driven by a density-wave order and lead to a clear sign change of the high-field Hall resistivity with temperature~\cite{stojkovic1996anomalous}. Despite its complex origin, a simplistic two-carrier model can still be used to extract valuable information on the density of hole-like and electron-like carriers in this system and give a fairly good estimate of their mobility as a function of temperature. In fact, a similar sign reversal of the Hall effect has been observed in several other hole-doped cuprates~\cite{leboeuf2007electron,doiron2013hall,leboeuf2011lifshitz,noda1999evidence,adachi2001crystal,suzuki2002hall}. As a novel perspective, it would be insightful to revisit the field dependence of the Hall resistivity in the vicinity of the crossing temperature of these materials to detect the nonlinear contribution.

Overall, the two-carrier model captures most of the features of the resistivity tensor involved in the description of the longitudinal and the Hall resistivity of electron-doped cuprates. It remains a very useful tool to explore the transport properties for doping above $x^{*}$ despite the fact that the Fermi surface is a featureless large hole-like cylinder and that the sign of the carriers is defined by strong electronic interactions. Our measurement of the nonlinear terms of the Hall resistivity and its agreement with the expectations from the two-carrier model represent a clear demonstration of its qualitative relevance even in this case as long as we assume that a mechanism driven by strong electronic interactions as that proposed by Kontani \textit{et al.}~\cite{kontani1999hall} is affecting the electron dynamics. Of course, the two-carrier model is a very naive one controlled by four parameters $n$, $p$, $\mu_n$ and $\mu_p$ that are difficult to relate directly to current-vertex corrections applied to a nearly antiferromagnetic Fermi liquid or to any other theory involving strong interactions. In a near future, it would be quite interesting to establish a direct link between these theories and the two-carriers model such that an analysis of the transport properties with the two-carrier model can be used in turn to better characterize the mechanism behind the unconventional behaviour of electronic excitations in cuprates.

\section{Conclusion}

In conclusion, we have measured the nonlinear Hall resistivity in electron-doped cuprates $\ce{Pr_{2-x}Ce_{x}CuO_{4 \pm \delta}}$ (\ce{PCCO}) as a function of doping $x$. The Hall resistivity measurements of \ce{PCCO} indicate that it is strongly temperature and doping dependent as previously reported. It shows sign changes in its temperature dependence in a narrow doping range ($ 0.16 \leq x \leq 0.18$) while it is entirely negative and positive in the extremely underdoped and overdoped region, respectively. The central focus of this paper is the presence of a nonlinear field dependence of the Hall resistivity close to the points in the phase diagram where the sign changes are observed. We show that the nonlinear contributions are present within the $0.125 \leq x \leq 0.20$ doping range in fact far away from the crossing points. The Hall resistivity, which includes a nonlinear field contribution, can be described using a two-carrier model where both hole-like and electron-like carriers contribute to the current density. The persistence of this nonlinear effect in the doping range where angle-resolved photoemission spectroscopy (ARPES) shows only a large hole-like cylindrical Fermi surface is a clear signal that strong electronic interactions affect the electron dynamics beyond $x^{*}$, the doping at which the pseudogap closes to lead way to only this large hole-like Fermi surface. Such observation agrees with a theoretical proposal that current-vertex corrections applied to nearly antiferromagnetic Fermi liquid are affecting electron dynamics. The negative sign of the nonlinear contribution to the Hall resistivity implies that the density of electron-like carriers is always larger than that of holes in the whole doping range of interest, up to $x = 0.20$. Moreover, the nonlinear term reaches a maximum for $x \approx x^{*}$ signalling a transition in the evolution of both densities with doping. A preliminary scenario is presented to account for the observations.\\

\section*{Acknowledgment}

The authors thank S. Pelletier, B. Rivard,  M. Abbasi Eskandari,  P. Reulet, E. Blais and F. Naud for technical support. The authors also gratefully acknowledge Prof. André-Marie Tremblay for helpful discussions.  This work is supported by the Natural Sciences and Engineering Research Council of Canada (NSERC) under grant RGPIN-2018-06656, the Canada First Research Excellence Fund (CFREF), the Fonds de Recherche du Québec - Nature et Technologies (FRQNT) and the Université de Sherbrooke.

\bibliographystyle{apsrev4-2}
\bibliography{reference.bib}
\end{document}